# Dynamic response of mesoscopic metal rings and thermodynamics at constant particle number

Jens Fricke and Peter Kopietz
*Institut für Theoretische Physik der Universität Göttingen,
Bunsenstraße 9, 37073 Göttingen, Germany*
(March 14, 1995)

## Abstract

We show by means of simple exact manipulations that the thermodynamic persistent current $I(\phi, N)$ in a mesoscopic metal ring threaded by a magnetic flux $\phi$ at constant particle number $N$ agrees *even beyond linear response* with the dynamic current $I_{dy}(\phi, N)$ that is defined via the response to a time-dependent flux in the limit that the frequency of the flux vanishes. However, it is impossible to express the disorder average of $I_{dy}(\phi, N)$ in terms of conventional Green's functions at flux-independent chemical potential, because the part of the dynamic response function that involves two retarded and two advanced Green's functions *is not negligible*. Therefore the dynamics *cannot* be used to map a canonical average onto a more tractable grand canonical one. We also calculate the zero frequency limit of the dynamic current at constant chemical potential beyond linear response and show that it is fundamentally different from any thermodynamic derivative.

Typeset using REVTEX

# I. INTRODUCTION

The existence of persistent currents in mesoscopic normal metal rings threaded by a magnetic flux $\phi$ has been postulated long time ago [1], although a clear theoretical understanding has been achieved only recently [2]. The experimental verification [3]-[5] of this effect has given rise to a renaissance of theoretical activity in this field [6]-[11]. Surprisingly, the magnitude of the measured average current was larger than the available theories could predict. The source for this discrepancy between theory and experiment remains controversial. One of us has proposed that the long-range nature of the Coulomb-interaction is responsible for the large observed current [11], but the issue remains controversial [12] and might eventually be settled numerically [13,14].

In the present work we shall put aside the fascinating problem of trying to understand the Coulomb-interaction, and add some new insights to the dynamic and thermodynamic aspects of non-interacting disordered electrons in an Aharonov-Bohm geometry. We consider disordered spinless electrons of mass $m$ on a three dimensional thin ring of circumference $L$ that is pierced by a flux $\phi$. The system can be described via the stochastic Hamiltonian

$$\hat{H} = \hat{H}_0 + U(\mathbf{r}) \quad , \quad \hat{H}_0 = \frac{\hat{\mathbf{P}}^2}{2m} \quad , \tag{1}$$

where the momentum operator is

$$\hat{\mathbf{P}} = \frac{\hbar}{i}\nabla_\mathbf{r} + \frac{2\pi\hbar\varphi}{L}\mathbf{e}_x \quad , \quad \varphi = \frac{\phi}{\phi_0} \quad . \tag{2}$$

Here $\phi_0 = hc/e$ is the flux quantum and $\mathbf{e}_x$ is a unit vector in azimuthal direction of the ring. Throughout this work the charge of the electron will be denoted by $-e$, and we call the azimuthal direction the $x$-direction. $U(\mathbf{r})$ is a Gaussian random potential with zero average and zero-range correlations,

$$\overline{U(\mathbf{r})} = 0 \tag{3}$$

$$\overline{U(\mathbf{r})U(\mathbf{r}')} = \tilde{\gamma}V\delta(\mathbf{r}-\mathbf{r}') \quad , \tag{4}$$

where $V$ is the volume of the ring. We follow here common praxis and assume only short range correlations between the random potential. It is possible, however, that an essential part of the physics that is responsible for the large observed currents has been lost within this approximation, see Ref. [15]. The parameter $\tilde{\gamma}$ can be related to the elastic lifetime $\tau$ by calculating the damping of the electrons to lowest order Born approximation. This leads to the identification

$$\tilde{\gamma} = \frac{\bar{\Delta}\hbar}{2\pi\tau} \quad , \tag{5}$$

where $\bar{\Delta}$ is the average level spacing at the Fermi energy. Throughout this work disorder averages will be denoted by an overbar. The level spacing $\Delta$ is defined via the relation

$$\Delta^{-1} = \frac{\partial N}{\partial \mu} = \sum_\alpha [-f'(\epsilon_\alpha - \mu)] \quad , \tag{6}$$

where the particle number $N$ as a function of the dimensionless flux $\varphi$ and the chemical potential $\mu$ is for a given realization of the disorder potential $U(\mathbf{r})$ given by

$$N(\varphi, \mu) = \sum_\alpha f(\epsilon_\alpha - \mu) \quad . \tag{7}$$

Here $f(\epsilon_\alpha - \mu)$ is the occupation of the exact energy level $\epsilon_\alpha$, and $f'(\epsilon)$ denotes the derivative of the function $f(\epsilon)$ with respect to its argument. In a canonical ensemble the function $f(\epsilon_\alpha - \mu)$ is not the Fermi function, although for $T \to 0$ it reduces to a step function irrespective of the choice of the ensemble. The energies $\epsilon_\alpha$ satisfy the single-particle Schrödinger equation

$$\hat{H}\psi_\alpha(\mathbf{r}) = \epsilon_\alpha \psi_\alpha(\mathbf{r}) \quad , \tag{8}$$

where the $\psi_\alpha(\mathbf{r})$ are the exact wave-functions. Because $\hat{H}$ depends parametrically on $\varphi = \phi/\phi_0$, the energies and wave-functions are functions of $\varphi$. Of course, in praxis it is impossible to calculate the $\epsilon_\alpha$ and $\psi_\alpha(\mathbf{r}) = <\mathbf{r}|\alpha>$ for fixed and arbitrary potential $U(\mathbf{r})$. Nevertheless, the disorder eigenbasis is useful for deriving exact relations between the various physical quantities.

It has been proposed by several authors [7]- [9] that the large measured currents might somehow be related to the fact that the rings in the experiment of Lévy et al. [3] were not attached to external leads. This implies that the number of electrons on each ring is fixed, so that for a proper theoretical description one should use a canonical ensemble [7]- [9]. One is then faced to the problem of calculating differences between grand canonical and canonical thermodynamic averages in mesoscopic systems. Unfortunately, there exist no systematic methods to calculate thermodynamic averages at constant particle number, and in praxis one has to use some kind of expansion in the fluctuations of the chemical potential [7]- [9]. A possible way to circumvent the averaging problem at constant particle number has been proposed by Efetov and collaborators [10,16–19]. Let us briefly outline the main ideas of this "dynamic approach" to the canonical averaging problem. Suppose that in addition to the static flux there is an oscillating component, $\phi(t) = \phi_0[\varphi + \int d\omega' e^{-i\omega' t/\hbar} \delta\varphi_{\omega'}]$. Note that in the experiment by Lévy et al. [3] the time dependent flux component is given by a sine-modulation with amplitude $\varphi_{ac} \approx \frac{1}{8}$ and frequency $\omega_0 \approx 10 - 10^3 Hz$, corresponding to $\delta\varphi_\omega = \frac{\varphi_{ac}}{2i}[\delta(\omega + \omega_0) - \delta(\omega - \omega_0)]$. Within linear response theory the time-dependent flux gives rise to a time dependent current $\delta I_{dy}(t, \varphi, \mu) = \int d\omega' e^{-i\omega' t/\hbar} \delta I_{dy}(\omega', \varphi, \mu)$ around the ring, with Fourier components given by

$$\delta I_{dy}(\omega, \varphi, \mu) = \frac{V\phi_0}{L^2}\left(\frac{e^2}{mcV}\right) K(\omega, \varphi, \mu)\delta\varphi_\omega + O(\delta\varphi_\omega^2) \quad , \tag{9}$$

where the so-called linear-response function $K(\omega, \varphi, \mu)$ is given by the Kubo-formula. We have rescaled the response function such that the prefactor in Eq.9 is given by

$$\frac{V\phi_0}{L^2}\left(\frac{e^2}{mcV}\right) = \frac{eh}{mL^2} = \left(\frac{ev_F}{L}\right)\frac{2\pi}{\Lambda} \quad , \quad \Lambda = k_F L \quad , \tag{10}$$

where $k_F$ is the Fermi wave-vector and $v_F = \hbar k_F/m$ is the Fermi velocity. For non-interacting electrons $K(\omega, \varphi, \mu)$ can then be written as

$$K(\omega,\varphi,\mu) = K^{dia}(\varphi,\mu) + K^{para}(\omega,\varphi,\mu) \quad , \qquad (11)$$

where the diamagnetic part is simply given by the negative particle number,

$$K^{dia}(\varphi,\mu) = -\sum_\alpha f(\epsilon_\alpha - \mu) = -N(\varphi,\mu) \quad , \qquad (12)$$

and the paramagnetic contribution is

$$K^{para}(\omega,\varphi,\mu) = \sum_{\alpha\beta} \frac{|P_{\alpha\beta}|^2}{m} \frac{f(\epsilon_\beta - \mu) - f(\epsilon_\alpha - \mu)}{\epsilon_\alpha - \epsilon_\beta - \omega - i0} \quad , \qquad (13)$$

with

$$P_{\alpha\beta} = \int d\mathbf{r}\, \psi_\alpha^*(\mathbf{r}) \hat{P}_x \psi_\beta(\mathbf{r}) \quad . \qquad (14)$$

Let us emphasize that Eq.11 is valid in a grand-canonical as well as in a canonical ensemble. Of course, in a grand canonical ensemble the electron number in Eq.12 depends on the disorder and the flux, while in the canonical ensemble we should solve Eq.7 to obtain the chemical potential as a function of $\varphi$ and $N$. We shall write $\mu(\varphi)$ for the fluctuating canonical chemical potential, and $\overline{\mu}_0$ for the constant grand canonical one. If an equation is valid for both ensembles provided the correct value for the chemical potential is substituted, we shall simply use the symbol $\mu$.

The fundamental assumption of the dynamic approach is [16,17]

$$\overline{K^{para}(0,\varphi,\overline{\mu}_0)} \approx \overline{K^{para}(0,\varphi,\mu(\varphi))} \quad . \qquad (15)$$

This equation tells us that the disorder average of the paramagnetic part of the response function is not very sensitive to the choice of the ensemble. Although this replacement is claimed to be valid for the paramagnetic part of the linear response function, in a canonical ensemble $K(0,\varphi,\mu(\varphi))$ is the derivative of a flux-periodic function with respect to the flux [16,17], so that the average over one period must vanish,

$$\int_0^1 d\varphi' K(0,\varphi',\mu(\varphi')) = 0 \quad . \qquad (16)$$

It follows that at constant $N$ we have exactly

$$N = \int_0^1 d\varphi' K^{para}(0,\varphi',\mu(\varphi')) \quad . \qquad (17)$$

Using Eqs.15 and 17 and dividing both sides of Eq.9 by $\delta\varphi_\omega$, one can write the average dynamic current in the limit $\omega \to 0$ as follows,

$$\lim_{\omega \to 0} \frac{\overline{\delta I_{dy}(\omega,\varphi,\mu(\varphi))}}{\delta\varphi_\omega} \approx \left(\frac{ev_F}{L}\right) \frac{2\pi}{\Lambda} \left[\overline{K^{para}(0,\varphi,\overline{\mu}_0)} - \int_0^1 d\varphi' \overline{K^{para}(0,\varphi',\overline{\mu}_0)}\right] \quad . \qquad (18)$$

Averaging at constant particle number has now been mapped onto the corresponding averaging at constant chemical potential, which can be performed by standard Green's function techniques.

There are some hidden but important subtleties that have been ignored so far. First of all, Eq.18 has been derived within linear response theory, which neglects the terms of higher order in $\delta\varphi_\omega$ in Eq.9. The interpretation of the left-hand side of Eq.18 as an ordinary derivative is therefore only justified if the non-linear terms that have been neglected in Eq.9 are consistent with the higher order flux-derivatives of the right-hand side of Eq.18. Moreover, even if this is the case, it is by no means obvious that the substitution in Eq.15 is indeed correct. At least one should be able to estimate the corrections to this replacement. In fact, in Ref. [17] Eq.18 was evaluated non-perturbatively in the diffusive regime, with the result that it does *not* precisely agree with the non-perturbative evaluation of an alternative approximate expression for $\overline{I(\varphi,\mu(\varphi))}$ [20]. This discrepancy has never been resolved. In this work we shall therefore critically re-examine the validity of Eqs.15 and 18. Our main results are (a) that the quadratic response is indeed consistent with the second flux-derivative of the equilibrium current at constant particle number, and (b) that the most crucial assumption 15 is not correct. The result (b) implies that the dynamics cannot be used to map a canonical averaging problem onto a grand canonical one, and in praxis one cannot avoid the expansion in powers of the fluctuations of the chemical potential [7]- [9], [21]. We derive the correct relation between dynamics and thermodynamics at constant particle number and explain discrepancies between various approaches that can be found in the literature. Finally, we shall discuss the grand canonical linear and quadratic dynamic response functions, which are *not* derivatives of a flux-periodic function. In the case of linear response we re-derive the non-perturbative results of Ref. [10] by means of a simple diagrammatic calculation.

## II. THE DYNAMIC CURRENT AND ITS RELATION TO THE CANONICAL EQUILIBRIUM CURRENT

The equilibrium persistent current is given by

$$I(\varphi,\mu) = \frac{(-e)}{mL}\sum_\alpha P_{\alpha\alpha} f(\epsilon_\alpha - \mu) \quad . \tag{19}$$

Using the Hellman-Feynman theorem, diagonal matrix elements of the momentum operator can be obtained from the flux-derivative of the exact eigen-energies $\epsilon_\alpha$,

$$\frac{\partial \epsilon_\alpha}{\partial \varphi} = \int d\mathbf{r}\, \psi_\alpha^*(\mathbf{r})\frac{\partial \hat{H}_0}{\partial \varphi}\psi_\alpha(\mathbf{r}) = \frac{2\pi\hbar}{L}\frac{P_{\alpha\alpha}}{m} \quad . \tag{20}$$

Because the three variables $\{\varphi,\mu,N\}$ are related via Eq.7, only two of them are independent, and we must specify the variable that is held constant if we differentiate with respect to $\varphi$. As usual in thermodynamics, the flux derivative of a thermodynamic quantity $X$ at constant $\mu$ will be denoted by $(\frac{\partial X}{\partial \varphi})_\mu$, while the corresponding derivative at constant $N$ is written as $(\frac{\partial X}{\partial \varphi})_N$. Note that the single-particle levels $\epsilon_\alpha$ are not thermodynamic quantities, so that the flux-derivative in Eq.20 does not depend on whether $\mu$ or $N$ is held constant. The following manipulations are valid for a given realization of the disorder potential and for a given particle number, so that subtleties associated with the averaging procedure [22] are irrelevant.

### A. Linear response

Taking the flux-derivative of Eq.20 at constant $\mu$, we obtain

$$\left(\frac{\partial I}{\partial \varphi}\right)_\mu = \frac{V\phi_0}{L^2}\left(\frac{e^2}{mcV}\right)\chi_{gc}(\varphi,\mu) \quad , \tag{21}$$

where the dimensionless grand canonical susceptibility is given by

$$\chi_{gc}(\varphi,\mu) = -\frac{L}{2\pi\hbar}\sum_\alpha\left[\frac{\partial P_{\alpha\alpha}}{\partial \varphi}f(\epsilon_\alpha - \mu) + P_{\alpha\alpha}\left(\frac{\partial f(\epsilon_\alpha - \mu)}{\partial \varphi}\right)_\mu\right] \quad , \tag{22}$$

with

$$\left(\frac{\partial f(\epsilon_\alpha - \mu)}{\partial \varphi}\right)_\mu = f'(\epsilon_\alpha - \mu)\frac{\partial \epsilon_\alpha}{\partial \varphi} = f'(\epsilon_\alpha - \mu)\frac{2\pi\hbar}{L}\frac{P_{\alpha\alpha}}{m} \quad . \tag{23}$$

Note that we have rescaled the susceptibility such that it can be directly compared with the linear response function $K_{dy}(0,\varphi,\mu)$ defined in Eqs.11-13. On the other hand, the flux-derivative at constant $N$ yields

$$\left(\frac{\partial I}{\partial \varphi}\right)_N = \frac{V\phi_0}{L^2}\left(\frac{e^2}{mcV}\right)\chi_c(\varphi,\mu) \quad , \tag{24}$$

where the dimensionless canonical susceptibility is

$$\chi_c(\varphi,\mu) = -\frac{L}{2\pi\hbar}\sum_\alpha \frac{\partial P_{\alpha\alpha}}{\partial \varphi}f(\epsilon_\alpha - \mu) \quad . \tag{25}$$

Note that in a canonical ensemble the derivative $\left(\frac{\partial f(\epsilon_\alpha - \mu)}{\partial \varphi}\right)_N$ must vanish [17]. Comparing Eqs.22 and 25, we conclude that

$$\chi_c(\varphi,\mu) = \chi_{gc}(\varphi,\mu) + \sum_\alpha \frac{P_{\alpha\alpha}^2}{m}f'(\epsilon_\alpha - \mu) \quad . \tag{26}$$

Let us now compare Eq.26 with the dynamic response function defined in Eq.11. In an infinite system, the dia- and paramagnetic contributions cancel exactly in the limit $\omega \to 0$ for any finite disorder. In a mesoscopic system, however, the cancellation is not perfect, so that in the static limit the dynamic susceptibility

$$\chi_{dy}(\varphi,\mu) = \lim_{\omega \to 0} K(\omega,\varphi,\mu) \tag{27}$$

is finite. In order to see the almost perfect cancellation between the dia- and paramagnetic parts of $\chi_{dy}(\varphi,\mu)$, we express Eqs.11-13 in terms of Green's functions. Defining the advanced and retarded Green's functions as usual,

$$G_\alpha^A(\epsilon) = \frac{1}{\epsilon - \epsilon_\alpha - i0} \quad , \quad G_\alpha^R(\epsilon) = \frac{1}{\epsilon - \epsilon_\alpha + i0} \quad , \tag{28}$$

we have exactly

$$K(\omega, \varphi, \mu) = -N(\varphi, \mu) + K_{eq}^{para}(\omega, \varphi, \mu) + K_{dy}^{para}(\omega, \varphi, \mu) \quad . \tag{29}$$

with

$$K_{eq}^{para}(\omega, \varphi, \mu) = \sum_{\alpha\beta} \frac{|P_{\alpha\beta}|^2}{m} \int_{-\infty}^{\infty} \frac{d\epsilon}{2\pi i} f(\epsilon - \mu) \left[ G_\alpha^R(\epsilon + \omega) G_\beta^R(\epsilon) - G_\alpha^A(\epsilon) G_\beta^A(\epsilon - \omega) \right] \tag{30}$$

and

$$K_{dy}^{para}(\omega, \varphi, \mu) = \sum_{\alpha\beta} \frac{|P_{\alpha\beta}|^2}{m} \int_{-\infty}^{\infty} \frac{d\epsilon}{2\pi i} \left[ f(\epsilon - \mu) - f(\epsilon - \omega - \mu) \right] G_\alpha^R(\epsilon) G_\beta^A(\epsilon - \omega) \quad . \tag{31}$$

Using the fact that

$$\frac{\partial |\alpha>}{\partial \varphi} = \frac{2\pi\hbar}{L} \sum_{\substack{\beta \\ (\beta \neq \alpha)}} \frac{P_{\beta\alpha}}{m} \frac{1}{\epsilon_\alpha - \epsilon_\beta} |\beta> \quad , \tag{32}$$

$$\frac{\partial P_{\alpha\alpha}}{\partial \varphi} = \frac{2\pi\hbar}{L} \left[ 1 + 2 \sum_{\substack{\beta \\ (\beta \neq \alpha)}} \frac{|P_{\alpha\beta}|^2}{m} \frac{1}{\epsilon_\alpha - \epsilon_\beta} \right] \quad , \tag{33}$$

it is easy to show that in the limit $\omega \to 0$ the first and second term in Eq.29 agree exactly with the function $\chi_{gc}(\varphi, \mu)$ given in Eq.22, i.e.

$$\chi_{gc}(\varphi, \mu) = -N(\varphi, \mu) + K_{eq}^{para}(0, \varphi, \mu) \quad . \tag{34}$$

Recall that in a finite disordered metal the energy levels repel each other [23], so that there are no degeneracies. As far as $K_{dy}^{para}(\omega, \varphi, \mu)$ is concerned, we note that the interval of integration in Eq.31 vanishes as $\omega \to 0$, but the terms with $\epsilon_\alpha = \epsilon_\beta$ in the sum give rise to a $1/\omega$-singularity. Cancelling this singularity against the factor of $\omega$ from the interval of integration, we obtain

$$\lim_{\omega \to 0} K_{dy}^{para}(\omega, \varphi, \mu) = \sum_{\alpha\beta} \delta_{\epsilon_\alpha, \epsilon_\beta} \frac{|P_{\alpha\beta}|^2}{m} f'(\epsilon_\alpha - \mu) \quad , \tag{35}$$

where $\delta_{\epsilon_\alpha, \epsilon_\beta}$ is unity if the discrete energy levels $\epsilon_\alpha$ and $\epsilon_\beta$ agree, and vanishes otherwise. We conclude that

$$\chi_{dy}(\varphi, \mu) = \chi_{gc}(\varphi, \mu) + \sum_{\alpha\beta} \delta_{\epsilon_\alpha, \epsilon_\beta} \frac{|P_{\alpha\beta}|^2}{m} f'(\epsilon_\alpha - \mu) \quad , \tag{36}$$

Comparing Eqs.26 and 36 and using the fact that $\delta_{\epsilon_\alpha, \epsilon_\beta} = \delta_{\alpha, \beta}$ because there are no degeneracies, we conclude that

$$\chi_{dy}(\varphi,\mu) = \chi_c(\varphi,\mu) \quad . \tag{37}$$

In spite of the fact that the above manipulations are very simple and exact, the dynamic current calculated in Ref. [17] in the diffusive regime by means of the supersymmetric $\sigma$-model [23] does not agree with the canonical equilibrium current that has been calculated in Refs. [7]- [9] via perturbation theory, and by Altland *et al.* [20] via the non-perturbative supersymmetry method. We now explain the origin of this discrepancy.

It is well-known in the theory of weak localization that disorder averages of products of Green's functions of the same type are "harmless" in the sense that they do not involve the singular contributions that arise in a perturbative approach due to the famous maximally crossed diagrams [24]. With this in mind, the authors of Refs. [10], [16]- [19] (which include one of the present authors) have not paid much attention to the disorder average $\overline{K_{eq}^{para}(0,\varphi,\mu(\varphi))}$, which involves the product of two retarded or two advanced Green's functions. Note that in the exact disorder-basis

$$K_{eq}^{para}(0,\varphi,\mu) = \sum_{\substack{\alpha\beta \\ (\epsilon_\alpha \neq \epsilon_\beta)}} \frac{|P_{\alpha\beta}|^2}{m} \frac{f(\epsilon_\beta - \mu) - f(\epsilon_\alpha - \mu)}{\epsilon_\alpha - \epsilon_\beta}$$
$$- \sum_{\alpha\beta} \delta_{\epsilon_\alpha,\epsilon_\beta} \frac{|P_{\alpha\beta}|^2}{m} f'(\epsilon_\alpha - \mu) \quad . \tag{38}$$

The second term in this expression cancels precisely the zero-frequency limit of $K_{dy}^{para}(\omega,\varphi,\mu)$ in Eq.35. Of course, if we work at constant chemical potential, then the disorder average of Eq.38 can be combined with the diamagnetic contribution $K^{dia}(\varphi,\mu) = -N(\varphi,\mu)$ to yield an exponentially small result after averaging, because these terms can be identified with the flux derivative of the grand canonical average equilibrium current. In this case the decomposition of $K^{para}$ into $K_{eq}^{para}$ and $K_{dy}^{para}$ is meaningful. However, in a canonical ensemble the first term in Eq.38 taken together with the diamagnetic contribution *is not exponentially small*, so that the above decomposition of $K^{para}$ is not useful. In other words, when evaluating the disorder average of $\chi_{dy}(\varphi,\mu(\varphi))$, it is not allowed to neglect the contribution from $\overline{-N(\varphi,\mu(\varphi))} + \overline{K_{eq}^{para}(0,\varphi,\mu(\varphi))}$.

To see how this can explain the discrepancies between the dynamic approach and Refs. [7]- [9], [20] consider the Fourier expansion of the equilibrium current $I(\varphi,\mu)$ in Eq.19, which is in general of the form

$$I(\varphi,\mu) = \sum_{n=1}^{\infty} I_n(\mu) \sin(2\pi n\varphi) \quad . \tag{39}$$

Here the Fourier coefficients $I_n(\mu)$ are functions of $\mu$ and functionals of the disorder potential. Similarly, we may expand the relation between $N, \mu$ and $\varphi$ in a Fourier series. Assuming that we have solved for $\mu$ as a function of $\varphi$ and $N$, we have

$$\mu(\varphi, N) = \sum_{n=0}^{\infty} \mu_n(N) \cos(2\pi n\varphi) \quad , \tag{40}$$

where the $\mu_n$ depend again on the disorder. Taking the derivative of Eq.39 with respect to $\varphi$ and setting then $\mu = \mu(\varphi)$, we obtain

$$\left(\frac{\partial I}{\partial \varphi}\right)_N = \sum_{n=1}^{\infty} I_n(\mu(\varphi)) 2\pi n \cos(2\pi n\varphi) + \sum_{n=1}^{\infty} \left.\frac{\partial I_n(\mu)}{\partial \mu}\right|_{\mu=\mu(\varphi)} \left(\frac{\partial \mu}{\partial \varphi}\right)_N \sin(2\pi n\varphi) \quad . \tag{41}$$

Evidently the first term on the right-hand side of this expression corresponds to the term $\chi_{gc}(\varphi, \mu(\varphi))$ in Eqs.26 and 36, which has been ignored in Refs. [10], [16]- [19]. The crucial point is now that *both terms in Eq.41 have the same order of magnitude*. Setting $\mu(\varphi) = \overline{\mu}_0 + \delta\mu(\varphi)$, where $\overline{\mu}_0$ is the disorder average of the zeroth Fourier component in Eq.40, and expanding

$$I_n(\overline{\mu}_0 + \delta\mu) \approx I_n(\overline{\mu}_0) + \left.\frac{\partial I_n(\mu)}{\partial \mu}\right|_{\mu=\overline{\mu}_0} \delta\mu + O(\delta\mu^2) \quad , \tag{42}$$

we have to leading order in $\delta\mu$,

$$\begin{aligned}\left(\frac{\partial I}{\partial \varphi}\right)_N &= \sum_{n=1}^{\infty} I_n(\overline{\mu}_0) 2\pi n \cos(2\pi n\varphi) \\ &+ \sum_{n=1}^{\infty} \left.\frac{\partial I_n(\mu)}{\partial \mu}\right|_{\mu=\overline{\mu}_0} 2\pi n \cos(2\pi n\varphi) \left[\mu_0 - \overline{\mu}_0 + \sum_{n'=1}^{\infty} \mu_{n'} \cos(2\pi n'\varphi)\right] \\ &- \sum_{n,n'=1}^{\infty} \left.\frac{\partial I_n(\mu)}{\partial \mu}\right|_{\mu=\overline{\mu}_0} 2\pi n' \mu_{n'} \sin(2\pi n\varphi) \sin(2\pi n'\varphi) \quad .\end{aligned} \tag{43}$$

Note that the terms with $n = n'$ in the double sums contain flux-independent contributions that would give rise to an aperiodic current. But these contributions cancel exactly if the second and third term are combined, as can be easily seen by writing

$$\begin{aligned}&n \cos(2\pi n\varphi) \cos(2\pi n'\varphi) - n' \sin(2\pi n\varphi) \sin(2\pi n'\varphi) \\ &= \frac{1}{2}\left[(n-n')\cos(2\pi(n-n')\varphi) + (n+n')\cos(2\pi(n+n')\varphi)\right] \quad .\end{aligned} \tag{44}$$

Therefore the average over a period of the right-hand side of Eq.44 indeed vanishes, in agreement with Eq.16. However, if only the last term in Eq.43 is retained, one obtains an aperiodic current that varies linearly with the flux [10]. The second term in 43 explains the differences between Ref. [17] and [20]. In the quasi-ballistic regime discussed in Ref. [16] the term with $\mu - \overline{\mu}_0$ in the second line of Eq.43 can be ignored after averaging, because disorder averages can be factorized. Moreover, the average current is dominated by the diagonal terms $n = n'$ in the double sums. Then it is easy to see that the second term is proportional to $\cos^2(2\pi n\varphi) = \frac{1}{2}[\cos(4\pi n\varphi) + 1]$, while the last term is proportional to $-\sin^2(2\pi n\varphi) = \frac{1}{2}[\cos(4\pi n\varphi) - 1]$. Obviously the constant terms cancel and the average current due to the last term is exactly half as large as the total current. Unfortunately, the factor of two due to the omission of the second term in Eq.43 was compensated in Ref. [16] by another mistake, so that the final result was correct and this discrepancy has not been noticed. We shall come back to this mistake in Sec.III. In the diffusive regime the averages of products cannot be factorized, so that the term containing $\mu_0 - \bar{\mu}_0$ is not negligible. Hence, even after subtraction of the flux-independent constant [16,17], the current calculated from the last term is in the diffusive regime not simply a factor of two smaller than the correct result, but its harmonic content is also different. At constant chemical potential the dynamic susceptibility is in general not the derivative of a periodic function of the flux. This can be seen from Eq.36, since $\chi_{gc}(\varphi,\mu)$ is the derivative of a flux-periodic function, but the term $\sum_{\alpha\beta} \delta_{\epsilon_\alpha,\epsilon_\beta} \frac{|P_{\alpha\beta}|^2}{m} f'(\epsilon_\alpha - \mu)$ is in general not. We shall discuss this term in more detail in Sec.III.

## B. Beyond linear response

So far, we have shown that in a canonical ensemble the flux-derivative of the thermodynamic persistent current agrees with the zero-frequency limit of the linear response kernel, i.e.

$$\lim_{\omega\to 0} \frac{\delta I_{dy}(\omega,\varphi,\mu(\varphi))}{\delta\varphi_\omega} = \left(\frac{\partial I}{\partial \varphi}\right)_N \quad . \tag{45}$$

This suggests that also the higher order flux-derivatives of the equilibrium current at constant $N$ agree with the static limit of the corresponding higher order dynamic response functions. We now show explicitly this agreement for the quadratic response. Writing

$$I_{dy} = \delta I_{dy} + \delta^2 I_{dy} + O(\delta\varphi^3) \tag{46}$$

the second order contribution to the dynamic current is

$$\delta^2 I_{dy}(t,\varphi,\mu) = \frac{1}{2}\left(\frac{ev_F}{L}\right)\frac{2\pi}{\Lambda}\int d\omega_1 d\omega_2 K^{(2)}(\omega_1,\omega_2,\varphi,\mu)e^{-i(\omega_1+\omega_2)t/\hbar}\delta\varphi_{\omega_1}\delta\varphi_{\omega_2} \tag{47}$$

with the quadratic response kernel given by

$$K^{(2)}(\omega_1,\omega_2,\varphi,\mu) =$$
$$-\left(\frac{2\pi\hbar}{L}\right)\left\{2\sum_{\alpha,\beta}\frac{P_{\alpha\beta}\delta_{\beta\alpha}}{m}\frac{f(\epsilon_\beta-\mu)-f(\epsilon_\alpha-\mu)}{\epsilon_\beta-\epsilon_\alpha+\omega_1+i0}\right.$$
$$+\sum_{\alpha,\beta}\frac{P_{\alpha\beta}\delta_{\beta\alpha}}{m}\frac{f(\epsilon_\beta-\mu)-f(\epsilon_\alpha-\mu)}{\epsilon_\beta-\epsilon_\alpha+\omega_1+\omega_2+i0}$$
$$+2\sum_{\alpha\beta\gamma}\frac{P_{\alpha\beta}P_{\beta\gamma}P_{\gamma\alpha}}{m^2}\frac{1}{\epsilon_\gamma-\epsilon_\alpha+\omega_1+\omega_2+i0}$$
$$\left.\times\left[-\frac{f(\epsilon_\beta-\mu)-f(\epsilon_\alpha-\mu)}{\epsilon_\beta-\epsilon_\alpha+\omega_1+i0}+\frac{f(\epsilon_\gamma-\mu)-f(\epsilon_\beta-\mu)}{\epsilon_\gamma-\epsilon_\beta+\omega_2+i0}\right]\right\} \quad . \tag{48}$$

This expression can be derived for example by non-equilibrium Green's function methods. It gives the stationary state after adiabatically switching on the periodic time-dependent flux components. The first and second term contain one current-vertex and one charge-vertex and vanish exactly. The first term accounts for the first order variation of the total number of electrons which, of course, is zero. The second term gives the variation of the electron distribution functions due to the term in the Hamiltonian which is quadratic in the time-dependent vector potential. As this term is proportional to the operator for the total number of electrons, it commutes with the Hamiltonian and gives no contribution in this dynamic problem. The quadratic response kernel is therefore given by the third term containing three current-vertices, which can be written as

$$K^{(2)}(\omega_1, \omega_2, \varphi, \mu) = -2 \left(\frac{2\pi\hbar}{L}\right) \sum_{\alpha\beta\gamma} \frac{P_{\alpha\beta} P_{\beta\gamma} P_{\gamma\alpha}}{m^2} \left\{ \frac{f(\epsilon_\alpha - \mu)}{(\epsilon_\alpha - \epsilon_\beta - \omega_1 - i0)(\epsilon_\alpha - \epsilon_\gamma - \omega_1 - \omega_2 - i0)} \right.$$
$$+ \frac{f(\epsilon_\beta - \mu)}{(\epsilon_\beta - \epsilon_\alpha + \omega_1 + i0)(\epsilon_\beta - \epsilon_\gamma - \omega_2 - i0)}$$
$$\left. + \frac{f(\epsilon_\gamma - \mu)}{(\epsilon_\gamma - \epsilon_\alpha + \omega_1 + \omega_2 + i0)(\epsilon_\gamma - \epsilon_\beta + \omega_2 + i0)} \right\} . \quad (49)$$

Defining the dynamic second-order susceptibility as the static limit of the second-order response kernel, i. e.

$$\chi^{(2)}_{dy}(\varphi, \mu) = \lim_{\substack{\omega_1 \to 0 \\ \omega_2 \to 0}} K^{(2)}(\omega_1, \omega_2, \varphi, \mu) \quad , \quad (50)$$

we obtain from Eq.49

$$\chi^{(2)}_{dy}(\varphi, \mu) = -2 \left(\frac{2\pi\hbar}{L}\right) \sum_{\alpha\beta\gamma} \frac{P_{\alpha\beta} P_{\beta\gamma} P_{\gamma\alpha}}{m^2} \quad (51)$$
$$\times \begin{cases} \frac{f(\epsilon_\alpha-\mu)}{(\epsilon_\alpha-\epsilon_\beta)(\epsilon_\alpha-\epsilon_\gamma)} + \frac{f(\epsilon_\beta-\mu)}{(\epsilon_\beta-\epsilon_\alpha)(\epsilon_\beta-\epsilon_\gamma)} + \frac{f(\epsilon_\gamma-\mu)}{(\epsilon_\gamma-\epsilon_\alpha)(\epsilon_\gamma-\epsilon_\beta)} & , \; \alpha \neq \beta \neq \gamma \neq \alpha \\ \frac{f(\epsilon_\gamma-\mu)-f(\epsilon_\alpha-\mu)}{(\epsilon_\alpha-\epsilon_\gamma)^2} & , \; \alpha = \beta \neq \gamma \quad \text{and cyclic permutations} \\ 0 & , \; \alpha = \beta = \gamma \end{cases} .$$

Rearranging by cyclic permutations we finally obtain

$$\chi^{(2)}_{dy}(\varphi, \mu) = -6 \left(\frac{2\pi\hbar}{L}\right) \left\{ {\sum_{\alpha\beta\gamma}}' \frac{P_{\alpha\beta} P_{\beta\gamma} P_{\gamma\alpha}}{m^2} \frac{f(\epsilon_\alpha - \mu)}{(\epsilon_\alpha - \epsilon_\beta)(\epsilon_\alpha - \epsilon_\gamma)} \right.$$
$$\left. + \sum_{\alpha \neq \gamma} \frac{P_{\alpha\alpha} |P_{\alpha\gamma}|^2}{m^2} \frac{f(\epsilon_\gamma - \mu) - f(\epsilon_\alpha - \mu)}{(\epsilon_\alpha - \epsilon_\gamma)^2} \right\} . \quad (52)$$

The prime at the first sum indicates that $\alpha \neq \beta \neq \gamma \neq \alpha$.

Now this result is compared with the canonical equilibrium second-order susceptibility defined by

$$\left(\frac{\partial^2 I}{\partial \varphi^2}\right)_N = \frac{e v_F}{L} \frac{2\pi}{\Lambda} \chi^{(2)}_c(\varphi, \mu) \quad . \quad (53)$$

In a canonical ensemble

$$\left(\frac{\partial^2 I}{\partial \varphi^2}\right)_N = \frac{-e}{mL} \sum_\alpha \frac{\partial^2 P_{\alpha\alpha}}{\partial \varphi^2} f(\epsilon_\alpha - \mu) \quad , \quad (54)$$

so that

$$\chi_c^{(2)}(\varphi,\mu) = -\frac{L}{2\pi\hbar}\sum_\alpha \frac{\partial^2 P_{\alpha\alpha}}{\partial\varphi^2} f(\epsilon_\alpha - \mu) \quad . \tag{55}$$

From second order time-independent perturbation theory for the non-degenerate case we have

$$\frac{\partial^2}{\partial\varphi^2}|\alpha> = \left(\frac{2\pi\hbar}{L}\right)^2 \frac{1}{m^2}\left\{-\sum_{\gamma(\neq\alpha)}\frac{|P_{\gamma\alpha}|^2}{(\epsilon_\alpha-\epsilon_\gamma)^2}|\alpha> \right. \tag{56}$$
$$\left. +2\sum_{\beta(\neq\alpha)}\left[\sum_{\gamma(\neq\alpha)}\frac{P_{\beta\gamma}P_{\gamma\alpha}}{(\epsilon_\alpha-\epsilon_\beta)(\epsilon_\alpha-\epsilon_\gamma)} - \frac{P_{\beta\alpha}P_{\alpha\alpha}}{(\epsilon_\alpha-\epsilon_\beta)^2}\right]|\beta>\right\} \quad ,$$

$$\frac{\partial^2 P_{\alpha\alpha}}{\partial\varphi^2} = 6\left(\frac{2\pi\hbar}{L}\right)^2\frac{1}{m^2}\left\{\sum_{\substack{\beta(\neq\alpha)\\\gamma(\neq\alpha)}}\frac{P_{\alpha\beta}P_{\beta\gamma}P_{\gamma\alpha}}{(\epsilon_\alpha-\epsilon_\beta)(\epsilon_\alpha-\epsilon_\gamma)} - \sum_{\gamma(\neq\alpha)}\frac{P_{\alpha\alpha}|P_{\alpha\gamma}|^2}{(\epsilon_\alpha-\epsilon_\gamma)^2}\right\} \quad . \tag{57}$$

We conclude that the second order canonical equilibrium susceptibility is given by

$$\chi_c^{(2)}(\varphi,\mu) = -6\left(\frac{2\pi\hbar}{L}\right)\frac{1}{m^2}\sum_\alpha f(\epsilon_\alpha-\mu)\left\{\sum_{\substack{\beta(\neq\alpha)\\\gamma(\neq\alpha)}}\frac{P_{\alpha\beta}P_{\beta\gamma}P_{\gamma\alpha}}{(\epsilon_\alpha-\epsilon_\beta)(\epsilon_\alpha-\epsilon_\gamma)} - \sum_{\gamma(\neq\alpha)}\frac{P_{\alpha\alpha}|P_{\alpha\gamma}|^2}{(\epsilon_\alpha-\epsilon_\gamma)^2}\right\} \quad . \tag{58}$$

This expression is easily seen to agree with the right-hand side of Eq. 52. Thus

$$\chi_{dy}^{(2)}(\varphi,\mu) = \chi_c^{(2)}(\varphi,\mu) \quad , \tag{59}$$

i. e. the equality of the canonical equilibrium susceptibility and the dynamic susceptibility holds in linear and quadratic order. We suspect that this agreement holds for all higher derivatives. We would like to emphasize, however, that we have not proven this agreement to all orders in perturbation theory.

### III. DYNAMICS AT CONSTANT CHEMICAL POTENTIAL

In this section we shall study the average dynamics of mesoscopic rings at constant chemical potential. Experimentally this corresponds to the dynamic response of an ensemble of rings that are somehow coupled to an external reservoir, so that the number of electrons can change as the flux is varied. Although it is hard to imagine how such a situation can be realized experimentally, the following calculation is instructive from the theoretical point of view, because it shows that the differences between canonical and grand canonical ensembles in mesoscopic rings manifest themselves in the dynamics perhaps even more drastically than in equilibrium properties.

## A. Linear response

We first discuss the linear response within a simple perturbative approach. Combining Eqs.34 and 36, setting $\mu = E_F = const$, and using $f'(\epsilon_\alpha - E_F) \to -\delta(\epsilon_\alpha - E_F)$ as $T \to 0$, we obtain in the zero-temperature limit for the average grand canonical dynamic susceptibility defined in Eq.27

$$\overline{\chi_{dy}(\varphi, E_F)} = -\overline{N(\varphi, E_F)} + \overline{K_{eq}^{para}(0, \varphi, E_F)} - \overline{\sum_{\alpha\beta} \delta_{\epsilon_\alpha,\epsilon_\beta} \frac{|P_{\alpha\beta}|^2}{m} \delta(\epsilon_\alpha - E_F)} \quad . \tag{60}$$

Let us now examine this expression in the diffusive regime, where the elastic mean free path $\ell$ is small compared with the circumference $L$ of the ring, but the localization length $\xi \approx M\ell$ is still larger than $L$. Here $M = \frac{(k_F L_\perp)^2}{4\pi}$ is the number of transverse channels, where $L_\perp$ is the transverse thickness of the ring. Due to an almost perfect cancellation between the dia- and paramagnetic contributions, the sum of the first two terms in Eq.60 is of order $e^{-L/\ell}$, i.e. exponentially small in the diffusive regime, so that only the last term survives. In Ref. [10] this term has been studied by means of the non-perturbative supersymmetry method, with the result that its average over the flux does not vanish. In subsequent work [16,17] this flux average was subtracted again, and not much attention was paid to this term. Because the machinery of the supersymmetry-method is physically not very transparent, let us give here a simple diagrammatic derivation of this contribution.

The last term on the right-hand side of Eq.60 involves a product of two matrix elements, but only a single energy denominator. Because the usual Green's functions in the momentum or real space basis combine always one matrix element with one energy denominator, such an expression cannot be directly written in terms of Green's functions. (Note that $\delta(\epsilon_\alpha - E_F) = \frac{1}{2\pi i}[G_\alpha^A(E_F) - G_\alpha^R(E_F)]$, so that a Dirac-$\delta$ should be counted as an energy denominator.) We therefore follow Ref. [16] and smooth out the Kronecker-$\delta$ by replacing

$$\overline{\sum_{\alpha\beta} \delta_{\epsilon_\alpha,\epsilon_\beta} |P_{\alpha\beta}|^2 \delta(\epsilon_\alpha - E_F)} \to \bar{\Delta} \overline{\sum_{\alpha\beta} P_{\alpha\alpha} P_{\beta\beta} \delta(\epsilon_\alpha - E_F) \delta(\epsilon_\beta - E_F)} \quad , \tag{61}$$

where the Jacobian for replacing the Kronecker-$\delta$ by the Dirac-$\delta$ is simply given by the average level spacing, see Eq.6. Note that for spinless electrons in three dimensions $\bar{\Delta} = \pi E_F/(\Lambda M)$, where $\Lambda = k_F L$. The Jacobian used in Eq.3.11 of Ref. [16] was too large by a factor of two, which led to an exact compensation of the mistake due to the omission of the second term in Eq.43, and the final result was correct.

Given Eq.61, the number of matrix elements matches again the number of the energy denominators, so that we may use standard Green's function techniques to average this expression. Going to the momentum basis $|\mathbf{k}>$, we have to calculate

$$\overline{\chi_{dy}(\varphi, E_F)} = -\bar{\Delta} \left(\frac{1}{2\pi i}\right)^2 \sum_{\mathbf{k}\mathbf{k}'} \frac{\hbar^2 \tilde{k}_x \tilde{k}'_x}{m} \overline{\left[G^A_{\mathbf{k},\mathbf{k}}(E_F) - G^R_{\mathbf{k},\mathbf{k}}(E_F)\right] \left[G^A_{\mathbf{k}',\mathbf{k}'}(E_F) - G^R_{\mathbf{k}',\mathbf{k}'}(E_F)\right]} \quad , \tag{62}$$

where $\tilde{k}_x = k_x + \frac{2\pi\varphi}{L}$, and

$$G^{A/R}_{\mathbf{k},\mathbf{k}'}(\epsilon) = \sum_\alpha \psi_\alpha(\mathbf{k})\psi^*_\alpha(\mathbf{k}')G^{A/R}_\alpha(\epsilon) \tag{63}$$

is the Green's function for a given realization of the disorder in the momentum basis. The wave-functions are $\psi_\alpha(\mathbf{k}) = <\mathbf{k}|\alpha>$. Because we are now working at fixed chemical potential, only the combination involving the product of an advanced and a retarded Green's function has to be retained in the diffusive regime. The dominant diagrams that determine the disorder average in Eq.62 are shown in Fig.1. These diagrams correspond to the following expression,

$$\overline{\chi_{dy}(\varphi,E_F)} = -\frac{2\bar\Delta}{(2\pi)^2}\sum_{\mathbf{k}\mathbf{k}'}\frac{\hbar^2 \tilde k_x \tilde k'_x}{m}\left[\overline{G}^A_\mathbf{k}(E_F)\right]^2\left[\overline{G}^R_{\mathbf{k}'}(E_F)\right]^2 [C(\mathbf{k}+\mathbf{k}',0) + D(\mathbf{k}-\mathbf{k}',0)] \quad, \tag{64}$$

where the averaged Green's functions are

$$\overline{G}^A_\mathbf{k}(\epsilon) = \frac{1}{\epsilon - \frac{\hbar^2 \mathbf{k}^2}{2m} - i\frac{\hbar}{2\tau}} \quad, \quad \overline{G}^R_\mathbf{k}(\epsilon) = \frac{1}{\epsilon - \frac{\hbar^2 \mathbf{k}^2}{2m} + i\frac{\hbar}{2\tau}} \quad, \tag{65}$$

and the Cooperon- and diffuson propagators are

$$C(\mathbf{q},\omega) = \tilde\gamma \frac{\frac{\hbar}{\tau}}{\hbar\mathcal{D}(\mathbf{q}+\frac{4\pi\varphi}{L}\mathbf{e}_x)^2 - i\omega} \quad. \tag{66}$$

$$D(\mathbf{q},\omega) = \tilde\gamma \frac{\frac{\hbar}{\tau}}{\hbar\mathcal{D}\mathbf{q}^2 - i\omega} \quad. \tag{67}$$

Here $\mathcal{D} = v_F\ell/3$ is the diffusion coefficient, and the parameter $\tilde\gamma$ is given in Eq.5. The diffuson contribution in Eq.64 has already been written down in Ref. [16], although it was not evaluated there. We would like to emphasize that the diffuson appears in Eq.64 on equal footing with the Cooperon. In contrast, the leading weak-localization correction to the Drude-formula involves only the Cooperon [24]. The reason for this difference is that the diagrams in Fig.1 are not conventional conductivity loops, but resemble more Hartree diagrams, which have the unusual feature that there is no energy integration associated with the Hartree loops. This is due to the fact that Eq.64 involves only propagators at the Fermi energy. From Eqs.66 and 67 it is clear that $D(0,0)$ and for $\varphi = 0$ also $C(0,0)$ are formally infinite. Within the framework of perturbation theory this singularity can be cured if contributions from higher order diagrams involving more diffusons and Cooperons are resummed [16]. Alternatively, we may simply use the results of the non-perturbative supersymmetric $\sigma$-model calculation [23], which imply that the correct way to regularize the singularity is to shift the frequency according to [25]

$$\omega \to \omega + i\Gamma \quad, \quad \Gamma = \frac{\bar\Delta}{\pi} \quad. \tag{68}$$

It is important to stress that terms with more than one Cooperon and diffuson are only important as far as the zero-mode is concerned. For all other modes the perturbative expansion in powers of Cooperons and diffusons is controlled by the small parameter $\frac{\bar\Delta}{E_c}$, where $E_c = \frac{\hbar\mathcal{D}}{L^2}$ is the Thouless energy. Note that $\frac{\bar{\Delta}}{E_c} = \frac{3\pi}{2}\frac{L}{M\ell} \propto \frac{L}{\xi}$, so that this expansion is good as long as the size of the system is small compared with the localization length $\xi$. In this regime it is sufficient to retain only the leading term involving one Cooperon and diffuson.

Using the regularization 68 and assuming that $L_\perp \lesssim \ell$, the evaluation of Eq.64 is straightforward. The final result is

$$\overline{\chi_{dy}(\varphi, E_F)} = -\frac{\Lambda}{(2\pi)^3 M} \sum_{n=-\infty}^{\infty} \left[\frac{1}{n^2 + \gamma} - \frac{1}{(n+2\varphi)^2 + \gamma}\right] \quad , \tag{69}$$

where we have defined

$$\gamma = \frac{\Gamma}{(2\pi)^2 E_c} = \frac{\bar{\Delta}}{4\pi^3 E_c} \quad . \tag{70}$$

The series in Eq.69 can be summed exactly. Using Eqs.9 and 10, we finally obtain for the dynamic current in the static limit

$$\lim_{\omega \to 0} \frac{\overline{\delta I_{dy}(\omega, \varphi, E_F)}}{\delta \varphi_\omega} = -\left(\frac{ev_F}{L}\right)\frac{1}{(2\pi)^2 M}\frac{\pi}{\sqrt{\gamma}}\left(\frac{1 + e^{-2\pi\sqrt{\gamma}}}{1 - e^{-2\pi\sqrt{\gamma}}}\right)$$
$$\times \left[1 - \frac{(1 - e^{-2\pi\sqrt{\gamma}})^2}{1 - 2\cos(4\pi\varphi)e^{-2\pi\sqrt{\gamma}} + e^{-4\pi\sqrt{\gamma}}}\right] \quad . \tag{71}$$

Expanding this expression for small $\gamma$ and using $\gamma^{-1} = \frac{2}{3}(2\pi)^2 M \frac{\ell}{L}$, we have to leading order

$$\lim_{\omega \to 0} \frac{\overline{\delta I_{dy}(\omega, \varphi, E_F)}}{\delta \varphi_\omega} = -\left(\frac{ev_F}{L}\right)\frac{2\ell}{3L}\left[1 - \frac{\bar{\Delta}}{\bar{\Delta} + 2\pi E_c[1 - \cos(4\pi\varphi)]}\right] \quad . \tag{72}$$

Note that the first term in the square brace is due to the diffuson pole, while the second term is due to the Cooperon. At zero flux both terms cancel, so that at $\varphi = 0$ the linear response function vanishes. For $|2\pi\varphi \bmod \pi| \gg \sqrt{\frac{\bar{\Delta}}{E_c}}$ the Cooperon contribution is completely negligible. Writing Eq.72 in terms of the diffusion coefficient $\mathcal{D} = v_F\ell/3$, we obtain in this regime

$$\lim_{\omega \to 0} \frac{\overline{\delta I_{dy}(\omega, \varphi, E_F)}}{\delta \varphi_\omega} = -\frac{2e\mathcal{D}}{L^2} \quad , \quad |2\pi\varphi \bmod \pi| \gg \sqrt{\frac{\bar{\Delta}}{E_c}} \quad . \tag{73}$$

This expression agrees exactly with Eq.18 of Ref. [10] (taking into account that we are working here with spinless electrons). Note that Efetov [10] has obtained this result by means of the non-perturbative supersymmetry-method, which is in principle exact. Therefore the approximations in Eqs.61 and 68 are justified a posteriori. In Ref. [17] the right hand side of Eq.72 is integrated over $\varphi$ after the average over the flux has been subtracted, and the result is interpreted as the canonical equilibrium current. From the analysis of Sec.II it is clear that such an interpretation is not correct. From our derivation it is also obvious that the value of the response function in the regime $|2\pi\varphi \bmod \pi| \gg \sqrt{\frac{\bar{\Delta}}{E_c}}$ is completely determined by the pole of the diffuson. Because the existence of this pole is a consequence of particle number conservation, Eq.72 should be not very sensitive to inelastic processes [10]. Note, however, that the zero-mode which is responsible for the large value of the linear term exists only in an isolated ring. If the current is measured via leads that are attached to the ring such that the azimuthal symmetry is broken, the zero mode should be omitted.

## B. Beyond linear response

We now calculate the leading non-linear correction to Eq.71. Writing Eq.48 in terms of Green's functions, we obtain

$$K^{(2)}(\omega_1, \omega_2, \varphi, E_F) = -\frac{2}{m^2}\left(\frac{2\pi\hbar}{L}\right)\sum_{\alpha\beta\gamma} P_{\alpha\beta}P_{\beta\gamma}P_{\gamma\alpha} \int_{-\infty}^{\infty} \frac{d\epsilon}{2\pi i}$$
$$\times \left\{[f(\epsilon - E_F) - f(\epsilon + \omega_2 - E_F)]G^R_\alpha(\epsilon + \omega_1 + \omega_2)G^R_\beta(\epsilon + \omega_2)G^A_\gamma(\epsilon)\right.$$
$$+[f(\epsilon + \omega_2 - E_F) - f(\epsilon + \omega_1 + \omega_2 - E_F)]G^R_\alpha(\epsilon + \omega_1 + \omega_2)G^A_\beta(\epsilon + \omega_2)G^A_\gamma(\epsilon)$$
$$+f(\epsilon - E_F)\left[G^A_\alpha(\epsilon)G^A_\beta(\epsilon - \omega_1)G^A_\gamma(\epsilon - \omega_1 - \omega_2)\right.$$
$$\left.\left.-G^R_\alpha(\epsilon + \omega_1 + \omega_2)G^R_\beta(\epsilon + \omega_2)G^R_\gamma(\epsilon)\right]\right\} \quad . \tag{74}$$

In a grand canonical ensemble we know that disorder averages involving the combinations $\overline{G^A G^A G^A}$ and $\overline{G^R G^R G^R}$ are exponentially small in the diffusive regime, so that only the first two terms in Eq.74 survive after averaging. We emphasize again that in the canonical ensemble this is not the case. The response function $K^{(2)}(-\omega, \omega, \varphi, E_F)$ in the frequency regime $\bar{\Delta} \lesssim \omega \lesssim E_c$ has been studied in Ref. [26]. Here we are interested in the limit $\omega_1, \omega_2 \to 0$, where we encounter the same problem as in the case of linear response: the number of matrix elements exceeds the number of energy denominators by one. In order to write this expression in terms of Green's functions, we introduce again an additional energy denominator via a smoothing procedure similar to the one used in Eq.61. In this way we obtain in the momentum basis

$$\overline{\chi^{(2)}_{dy}(\varphi, E_F)} = -3\bar{\Delta}\frac{\hbar^4}{m^2}\frac{2\pi}{L}\frac{1}{(2\pi i)^2} \sum_{\mathbf{k}\mathbf{k}'\mathbf{k}''} \tilde{k}_x \tilde{k}'_x \tilde{k}''_x$$
$$\times \overline{\left[G^A_{\mathbf{k},\mathbf{k}}(E_F) - G^R_{\mathbf{k},\mathbf{k}}(E_F)\right]\left[G^A_{\mathbf{k}',\mathbf{k}''}(E_F)G^A_{\mathbf{k}'',\mathbf{k}'}(E_F) - G^R_{\mathbf{k}',\mathbf{k}''}(E_F)G^R_{\mathbf{k}'',\mathbf{k}'}(E_F)\right]} \quad . \tag{75}$$

The prefactor of 3 is due to the three cyclic permutations in Eq.51. On the other hand, if we differentiate the average linear response function $\overline{\chi_{dy}(\varphi, E_F)}$ in Eq.62 with respect to $\varphi$, using the exact identity

$$\frac{\partial}{\partial \varphi} G^{A/R}_{\mathbf{k},\mathbf{k}'}(\epsilon) = \frac{\hbar^2}{m}\frac{2\pi}{L} \sum_{\mathbf{k}''} G^{A/R}_{\mathbf{k},\mathbf{k}''}(\epsilon)\tilde{k}''_x G^{A/R}_{\mathbf{k}'',\mathbf{k}'}(\epsilon) \quad , \tag{76}$$

we obtain an expression almost identical to Eq.75, the difference being that the prefactor of 3 is replaced by a prefactor of 2. We conclude that in the diffusive regime

$$\overline{\chi^{(2)}_{dy}(\varphi, E_F)} = \frac{3}{2}\frac{\partial}{\partial \varphi}\overline{\chi_{dy}(\varphi, E_F)} \quad . \tag{77}$$

Note that the factor of $\frac{3}{2}$ does not appear in the corresponding equation for the canonical ensemble, see Eq.59. According to Eq.77 the large coefficient of the linear response term in Eq.73 does not appear in the quadratic response, so that outside the narrow regions where $2\varphi$ is close to an integer the leading correction to the linear response is small, even if $\delta\varphi_\omega$ is of the order of unity.

## IV. CONCLUSIONS

In this work we have clarified the relation between the dynamic current and the thermodynamic current at constant particle number, and have settled a controversy that has remained unresolved for the past three years. We have shown that the fundamental assumption of the dynamic approach, Eq.15, is not justified, because the flux dependence of the chemical potential in the disorder average $\overline{K_{eq}^{para}(0,\varphi,\mu(\varphi))}$ cannot be neglected. Although this part of the response function involves only disorder averages of the form $\overline{G^A G^A}$ and $\overline{G^R G^R}$, its contribution to the average canonical persistent current is of the same order of magnitude as the contribution from the $\overline{G^R G^A}$-term that is usually retained. Thus, one of the most basic properties of disorder averages of products of Green's functions in the diffusive regime does not apply to mesoscopic systems at constant particle number. Hence, also in the calculation of dynamic properties one cannot avoid the problem of expanding in powers of the fluctuations of the chemical potential, assuming that such an expansion is allowed. In the light of this result previous calculations should be critically re-examined.

We have also calculated the leading corrections to the linear response functions and have shown that in a canonical ensemble the zero-frequency limit of the quadratic response can be obtained from the flux-derivative of the corresponding linear response function. In a grand-canonical ensemble, however, this is not the case. Finally, we have shown that non-perturbative results obtained via the supersymmetric $\sigma$-model can be exactly reproduced by combining diagrammatic perturbation theory with a simple regularization prescription of the Cooperon and diffuson propagators.

## ACKNOWLEDGEMENTS

We would like to thank K. Schönhammer for discussions and S. Kettemann for communications and for sending us a copy of his dissertation. We also appreciate comments by K. B. Efetov. This work was financially supported by the Deutsche Forschungsgemeinschaft (SFB 345 "Festkörper weit weg vom Gleichgewicht").

# FIGURES

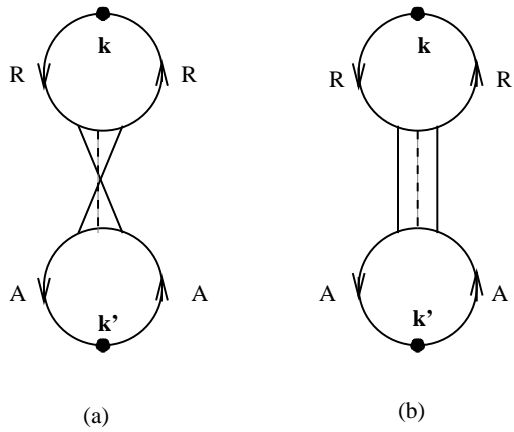

Fig.1
Fricke, Phys. Rev. B

FIG. 1. Dominant diagrams that determine the average dynamic susceptibility $\overline{\chi_{dy}(\varphi, E_F)}$, see Eq.64. A solid arrow with label $R$ or $A$ denotes an averaged retarded or advanced Green's function *with energy fixed at $E_F$*. The vertex in (a) represents the Cooperon, and the vertex in (b) represents the diffuson. The small black circles denote current vertices. It is understood that there is no energy integration associated with the loops.